\documentclass[conference]{IEEEtran}
\ifCLASSINFOpdf
\else
\fi
\usepackage{balance}       
\usepackage{graphics}      
\usepackage[T1]{fontenc}   
\usepackage{txfonts}
\usepackage{mathptmx}
\usepackage{hyperref}

\usepackage{color}
\usepackage{booktabs}
\usepackage{textcomp}

\usepackage{microtype}        
\usepackage{ccicons}          
\usepackage{multirow}
\usepackage{todonotes}
\usepackage{pifont}
\usepackage{tabu}

\newcommand{\para}[1]{{\vspace{4pt} \bf \noindent #1
\hspace{10pt}}}

\makeatletter
\def\url@leostyle{%
    \@ifundefined{selectfont}{
    \def\UrlFont{\sf}
    }{
    \def\UrlFont{\small\bf\ttfamily}
    }}
\makeatother
\def\pprw{8.5in}
\def\pprh{11in}

\DeclareRobustCommand{\cmark}{\ding{51}}%
\DeclareRobustCommand{\xmark}{\ding{53}}%

\begin{document}


\title{Towards the Adoption of Anti-spoofing Protocols}
\author{Hang Hu, Peng Peng, Gang Wang\\
Department of Computer Science, Virginia Tech\\
{\small \{hanghu, pengp17, gangwang\}@vt.edu}
}


\maketitle

\begin{abstract}
Email spoofing is a critical step in phishing attacks, where the attacker impersonates someone that the victim knows or trusts. In this paper, we conduct a qualitative study to explore why email spoofing is still possible after years of efforts to design, develop, and promote anti-spoofing protocols (SPF, DKIM, DMARC). 
Previous research shows that the adoption rates of anti-spoofing protocols are still very low.
To understand the reasons behind the slow adoption, we  conduct a user study 
with 9 email administrators from different institutions. The results show that email administrators are aware of the weaknesses of these protocols and believe the current protocol adoption lacks the crucial mass due to the protocol defects, weak incentives, and practical deployment challenges. Based on these results, we discuss the key implications to protocol designers, email providers and users, and future research directions to mitigate the email spoofing threats.

\end{abstract}



\section{Introduction}
Phishing attack is a known threat to the Internet. Recently, this threat has
been significantly escalated due to its heavy involvement in massive data breaches~\cite{verizon17}, ransomware outbreaks~\cite{ransom1}, and even political campaigns~\cite{DNC1}. For example, spear phishing emails have been used in nearly half of the 2000 reported breaches in 2016, responsible for leaking billions of data records~\cite{verizon17}.

Email spoofing is a critical step in phishing attacks where the
attacker impersonates someone that the victim knows or trusts. By
spoofing the email address of a reputable organization or
a close friend, the attacker has a better chance to deceive the victim~\cite{Jagatic:2007}.
To prevent spoofing, there has been an active effort since the early
2000 to develop, promote, and deploy anti-spoofing protocols. Protocols such
as SPF~\cite{spf}, DKIM~\cite{dkim}, and DMARC~\cite{dmarc} have
become the Internet standards, allowing email receivers to verify the
sender's identity. 

\begin{figure}[t]
\centering
\begin{minipage}{0.47\textwidth}
\includegraphics[width=0.98\textwidth]{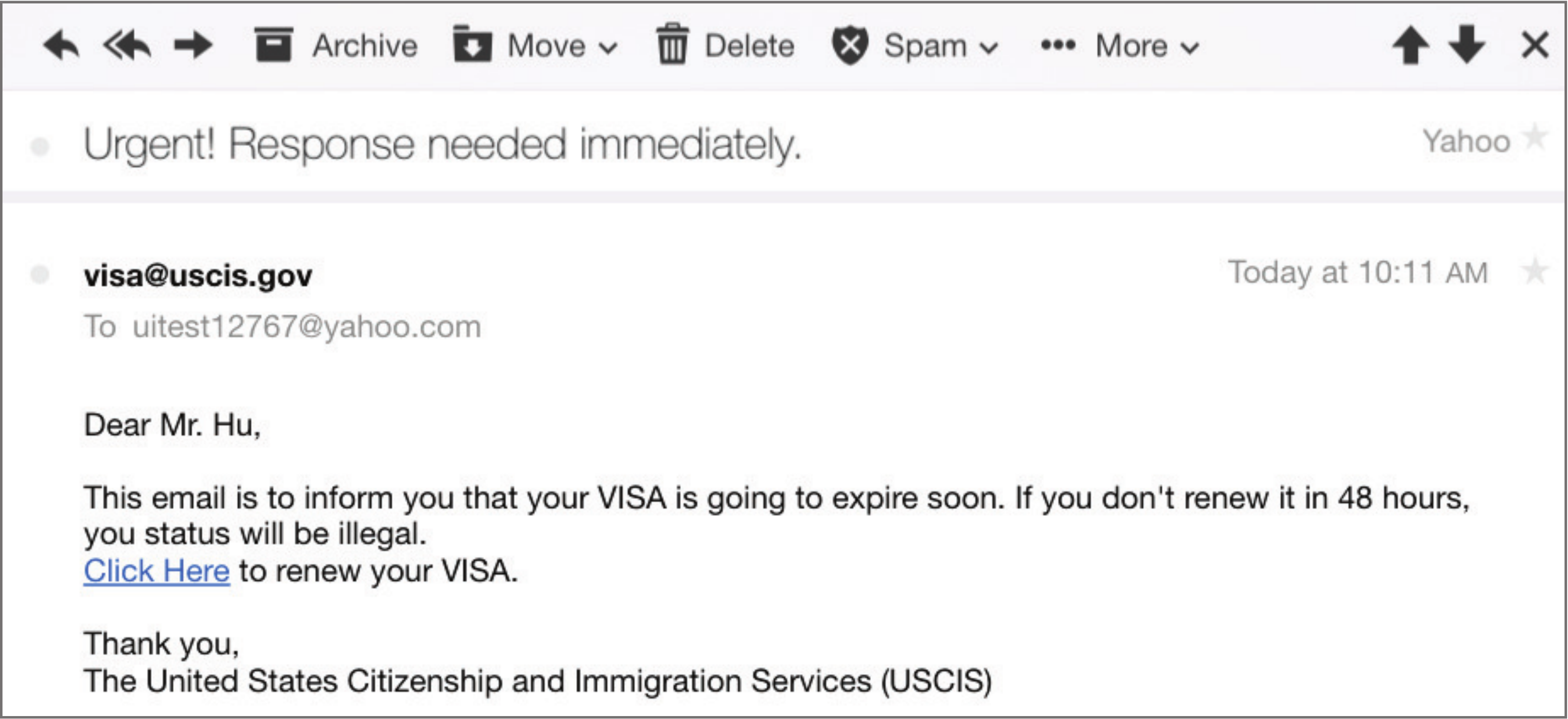}
\caption{A spoofing email that impersonates the U.S. Citizenship and
  Immigration Services (USCIS). We acted as the attacker and sent this
  email to our own account. The email arrived the inbox without
  triggering any alert (Mr. Hu is a fictional name).}
\label{fig:uscis}
\end{minipage}
\end{figure}

Despite these efforts, however, sending spoofing emails is still surprisingly
easy today. As an example, Figure~\ref{fig:uscis} shows a spoofing email where the
sender address is set to the domain of the U.S. Citizenship and Immigration Services
(USCIS). We crafted and sent this email to our own account in Yahoo (as the
victim), and it successfully reached the inbox without triggering any
warnings. This is not a coincident as email spoofing is still widely
used in real-world phishing attacks~\cite{verizon17, ransom3, DNC1}.

The real question is, {\em why email spoofing is still possible} after
years of efforts spent on the defense. In 2015, two measurement
studies~\cite{Durumeric:2015,Foster:2015}  show that the adoption rates of
anti-spoofing protocols are still low. Among Alexa top 1 million domains, only
40\% have adopted SPF and only 1\% have DMARC. 

In this paper, we perform a qualitative study to understand why anti-spoofing protocols are not widely adopted, particularly from email providers' perspectives. 
To understand the perception of email providers, we
conduct a user study with email administrators from different
institutions. This user study turned out to be challenging to
conduct. Part of the reason is that the candidate pool is small. People
who can provide insights for our questions need to have extensive
experience managing real-world email services. In addition, email
administrators often hesitate (or are not allowed) to share details about their
anti-phishing/spoofing solutions. To these ends, we send our user study requests to
4000 email administrators of Alexa top domains. We eventually
received responses from $N=9$ administrators from various
organizations (universities, payment services, online community websites)
who agree to answer open questions either online or
through in-person interviews. 

Our results show that email administrators are aware of the
technical weaknesses of SPF, DKIM and DMARC. The general perception is
that these protocols are ``helpful'', but ``cannot solve the spoofing
problem completely''.  The email administrators believe that the slow
adoption of the protocols is primarily due to the lack of a critical
mass. Like many network protocols, the benefits of the anti-spoofing
protocols come into existence only if a large number of Internet domains start to
adopt the protocols to publish their authentication
records. Currently, the incentive of adoption is not strong,
especially for Internet domains that don't host emails services (but they
still can be spoofed). In addition to the technical weaknesses, the
email administrators pointed out the practical challenges to deploy
the protocols, particularly, for organizations that use cloud-based
email services and large organizations that have many dependent
services. Finally, email administrators share their thoughts on the
possible solutions moving forward. One interesting direction is to
improve the current email user interface to support security
indicators, and educate users to proactively check email
authentication results.
  
In summary, our work makes three contributions. 
\begin{itemize}

     \item First, through the user study, we provide new
       insights into the perceived values and concerns of anti-spoofing
       protocols from email providers' perspectives. These results
       shed light to the reasons behind the 
       slow adoption of SPF, DKIM, and DMARC, pointing out the
       directions of improvement moving forward. 

      \item Second, we discuss the key implication of the results to
        protocol designers, email providers, and users. We discuss
        the possible solutions at the user-end to make up for the
        defective server-side authentication. 
\end{itemize}

\section{Background and Related Work}




In the following, we describe the background of email spoofing attacks
and anti-spoofing protocols. Then, we introduce related {\em technology adoption
  theories} to set up the contexts for our study. 

\subsection{SMTP and Email Spoofing}
Simple Mail Transfer Protocol (SMTP) is the Internet standard for
email transmission~\cite{smtp}. 
A key limitation of SMTP
is that it has no built-in security features to prevent people
(attackers) from impersonating/spoofing an arbitrary sender address. To perform a spoofing attack, attackers can manipulate two key fields
to send emails. First, after establishing an SMTP connection to the target mail server,
the attacker can use the ``{\tt MAIL FROM}'' command and set the sender
address to anyone that they want to impersonate. After that, the
``{\tt MAIL FROM}'' address is inserted into the header as the
``Return-Path''. In addition, attackers can modify another field called
``{\tt From}'' in the email header. This ``{\tt From}'' field specifies the address that will be displayed on the email
interface~\cite{email_message}. When a user receives the
email, the user will see the ``{\tt From}'' address ({\em e.g.}, {\em visa@uscis.gov}  in
Figure~\ref{fig:uscis}). If the user replies the email, the reply
message will go to the ``Return-Path'' set by ``{\tt MAIL FROM}''.
Note that the two addresses are not necessarily the same. 

Email spoofing is a critical step of phishing attacks to gain the victim's
trust~\cite{social-phishing, Hong:2012, Jagatic:2007}. Meanwhile,
spoofing is also a strong signal of phishing
attacks~\cite{Hong:2012,Krammer:2006:,Kumaraguru:2007:,McGrath:2008:,Prakash:2010:}. 
Spoofing detection results are often used by phishing
detection systems~\cite{DBLP:conf/ecrime,
  spear-phishing1, Fette:2007:L, 203674, google17}.




\subsection{Anti-Spoofing Protocols}
To detect and prevent email spoofing, SMTP extension protocols are
proposed including SPF, DKIM and DMARC. All three
protocols have been published or standardized by the Internet Engineering Task
Force (IETF). 

\para{SPF.} Sender Policy Framework (SPF) was proposed in
early 2000, and standardized in 2014~\cite{spf}. SPF
allows a domain to publish a list of IPs that are authorized to send
emails on its behalf. For instance, the domain  {\tt a.com} can
publish its SPF record in the DNS. When the receiving server receives
the {\tt MAIL FROM} command claiming to be
{\tt alex@a.com}, the receiving server can check if the sender IP is
listed in the SPF record of {\tt a.com}. 


\para{DKIM.} DomainKeys Identified Mail (DKIM) was first drafted in
2004 and standardized in 2011~\cite{dkim}. DKIM uses a public-key based approach
to authenticate the email sender and check the email
integrity. More specifically, the sender's email service will place a digital signature in
the email header signed by the private key associated with the sender's domain.
The receiving service can retrieve the sender's public key from DNS to verify the signature. 
To retrieve a DKIM public key from DNS, one will need
the selector information (an attribute in the DKIM signature beside the domain name. 
By verifying the DKIM signature, the receiver can detect
if the signed message has been modified, to ensure integrity and
authenticity.



\para{DMARC.}
Domain-based Message Authentication, Reporting and Conformance (DMARC) 
was drafted in 2011 and published in 2015~\cite{dmarc}. DMARC is not a standalone protocol
but needs to work with SPF and/or DKIM. DMARC allows the domain owner to publish a ``failing policy'' 
which specifies what actions the receiver should take 
when the incoming email fails the DMARC checks.
In addition, DMARC requires {\em identifier alignment} from SPF or
DKIM. For SPF, alignment means that {\tt MAIL FROM} address used for
the SPF check should be consistent with the {\tt From} field in
the header. For DKIM, alignment means that the domain name in the DKIM
signature should match the {\tt From} field. Alignment ensures the email
address that user sees matches with the authenticated address.

\subsection{Technology Adoption Theories}
To provide the contexts for our study, we briefly introduce the
important theories of technology adoptions~\cite{parente1994barriers, Lindley:2017:,
  katz1986technology, Kim:2015:EBA}. With a focus on the networking and
security protocols, prior works have examined the
adoption of DNSSEC~\cite{ching-imc-2017}, IPv6~\cite{Colitti:2010:}, 
HTTPS~\cite{203662}, Bitcoin~\cite{bohme2013internet},
and Biometric tracking~\cite{Ahmed:2017}. Below, we discuss three
adoption theories related to our study. 

\para{TAM.} 
Technology Acceptance Model (TAM) is the most basic theory that
models user intention to adopt new technologies~\cite{doi:10.1287,
  DECI:DECI192}. 
The model describes the key factors
that influence user decision, the most important of which are {\em
  perceived usefulness} (PU) and {\em perceived ease-of-use}
(PEOU). In our context, ``users'' refer to email services.
TAM has many extended versions with more
factors added to the model ({\em e.g.}, self-efficacy, quality of the system)~\cite{LAI2017, doi:10.1287, doi:10.1586,
  my2009overview}.

\begin{table*}[t]
\centering
\caption{User study participants: 9 email administrators. U8 requested
  to conceal the institution type, and thus we keep it as
  ``anonymous''. For each of their email services, we also measured
  whether the email domain published the DNS authentication records
  (as the sender) and whether the domain authenticate incoming emails (as the receiver).  
  }
\label{tab:institutions}
\begin{tabu}{|l|l|l|l|l|l|l|l|l|}
\tabucline[1.1pt]{-}
\multirow{2}{*}{UserID} & \multirow{2}{*}{User Study Method} &
\multirow{2}{*}{Email Service Type} & \multicolumn{3}{c|}{As Sender: Publish Records?} & \multicolumn{3}{c|}{As Receiver: Authenticate?} \\
\cline{4-9}
    &                       &                                     & SPF    & DKIM    & DMARC    & SPF    & DKIM   & DMARC   \\
\hline
U1  & In-person Interview   & University1 (campus-level)          & \cmark & /   & \cmark & \cmark & \cmark & \cmark  \\ \hline
U2  & In-person Interview   & University1 (department-level)      & \xmark   & /   & \xmark   & \cmark & \cmark & \cmark  \\ \hline
U3  & Open-question Survey  & Payment System                      & \cmark & /   & \cmark  & \cmark & \cmark & \cmark  \\\hline
U4  & Open-question Survey  & Website Hosting Service             & \cmark & /   & \xmark   & \xmark & \cmark & \xmark  \\\hline
U5  & Open-question Survey  & Advertisement Service1              & \cmark & /   & \cmark & \cmark & \cmark & \cmark  \\\hline
U6  & Open-question Survey  & Advertisement Service2              & \cmark & /   & \xmark   & \cmark & \cmark & \cmark  \\\hline
U7  & Open-question Survey  & University2 (campus-level)          & \xmark   & /   & \xmark   & \cmark & \cmark & \cmark \\\hline
U8  & Open-question Survey  & Anonymous                           & / & /   & /     & /    & /    & /    \\\hline
U9  & Open-question Survey  & Online Community                    & \cmark & /  & \xmark   & \cmark & \cmark & \cmark \\
\tabucline[1.1pt]{-}     
\end{tabu}
\end{table*}


\para{Network Externalities.} 
For network protocols, the standard TAM is usually not
sufficient to explain their adoption since individual user's
decision is likely to affect other users. This leads to
the notion of {\em Network Externalities} (or net effect)~\cite{katz1986technology, bohme2013internet}. Network
externalities mean that an individual adopter can add the value for
other people to adopt the same technology. In other words, when more users
adopt the same protocol, the value of the protocol to each
user will also increase~\cite{citeulike:126680}. For anti-spoofing
protocols, if more domains publish their SPF/DKIM/DMARC records, it makes
easier for other email providers to detect spoofing emails.


\para{Cost-Benefit Model.}
Ozment and Schechter propose an adoption model that focuses on the
cost-benefit perspective~\cite{ozment2006boot}. The
model argues that only when the {\em
  benefits} to individual adopters overweight the adoption {\em costs}
will the protocol be widely accepted. 
For network protocols, the per-user benefits may grow as more users adopt the protocol (net
effect)~\cite{aboba2008makes}. The costs can be either constant or
changing (mostly decreasing) as more users adopt the protocol. 

Often cases, a network protocol requires a minimal
level of deployment before creating enough benefits to overweight the
costs. This leading to notions of {\em critical mass} which
represents the minimal number of adopters in order to facilitate
self-sustaining adoption or create further growth~\cite{citeulike:126680}.


\section{Research Questions and Methodology}
In this paper, we conduct an exploratory study to understand the adoption of
anti-spoofing protocols. We qualitatively look into two key
questions. First, what are the reasons behind the relatively low
rate of anti-spoofing protocols? Second, why did most domain owners
configure the protocol with relaxed failing policies?

To answer these questions, 
we seek to understand the user perception of these
protocols in terms of the perceived usefulness (PU) and perceived
ease-of-use (PEOU) (two key aspects of TAM). Here, ``users'' does not
refer to the users of the email system. Instead, ``users'' refer to the email service administrators who will use the
protocol to defend against spoofing attacks. 
To understand email providers' perceptions towards the anti-spoofing
protocols, we conduct a user study (with IRB approval). The biggest
challenge for this user study is to recruit participants to share
their experience and insights. More specifically, we need to recruit
participants who have real-world experience of operating an email service
and/or deploying anti-spoofing protocols. This narrows down the
candidate pool to a small and highly specialized user population. In
addition, real-world email administrators are often
reluctant to share due to the sensitivity of the topic. For many
companies and organizations, details about their phishing/spoofing
detection systems are non-disclosable.

To address these challenges, we sent our user study requests to a
large number of email administrators. More specifically, we contacted
the email administrators of Alexa top 4000 domains. In the user study request,
we ask about their preferred ways of participation ({\em e.g.}, survey,
phone interviews) and the level of details they feel comfortable to
share. In total, we recruit $N=9$ email providers from different organizations. 7 participants agree to fill in a survey
with ``open questions'' and 2 participants agree to do an
in-person interview. In Table~\ref{tab:institutions}, we list the 9
email administrators and the {\em type} of their institutions and
organizations. Note that {\tt U8} requested to conceal the
institution-specific information, and thus we keep it as ``anonymous''. This small-scale but in-depth user study seeks to
provide useful qualitative results and new insights from {\em protocol users}' perspectives.

To provide the context for each email service that the participant
manages, we also performed a quick measurement as shown in
Table~\ref{tab:institutions}. We measured whether the email domain published the corresponding authentication records in DNS (as the
sender) and whether the domain performed authentication checks on the incoming
emails (as the receiver). Same as before, we cannot measure whether an
email domain has published the DKIM public key without knowing its selector
(marked with ``/''). We observe that most of the email services perform all three authentication checks on incoming
emails (7 out of 8) and one email service checks DKIM only. However,
when acting as the sender domain, only 3 email services published both SPF and DMARC records to the DNS.

For the interview and survey participants, we use the same list of open questions. The difference is that we can ask follow-up
questions to the interview participants, but not the survey
participants. Some of the detailed questions are designed based on the results
of {\em step1}, which we will discuss later. At the high-level, the
open questions fall into the following themes. First, we ask the
participants to comment on the email spoofing problem and how they
usually detect spoofing attempts. Second, we ask the participants to
comment on the value that SPF, DKIM and DMARC could bring in for their email
services. Third, we ask about their personal perceptions towards the
under-adoption of anti-spoofing protocols and the possible
reasons. Fourth, we ask why most of the deployed protocols were not configured
``strictly''. Finally, we ask the participants to comment on the
possible solutions moving forward to the email spoofing problem.  

The survey participants answer the open questions using an online
survey website that we set up. The interview participants then have a
face-to-face interview session for 45 to 
60 minutes. Our study is approved by IRB. We ensure that all the data are
properly anonymized and securely stored.

\section{Result: User Study}
Our user study focuses on open questions regarding the values and
concerns of SPF, DKIM and DMARC, and the possible reasons behind their slow
adoption. In the following, we discuss our findings by grouping the
results into 6 high-level topics. 

\subsection{Technical Defects of the Protocols}
Email administrators have acknowledged the values of adoption
these protocols. However, the most discussed topics are still the technical flaws in SPF, DKIM and DMARC. We summarize these tehnical flaws as the following.

First, SPF and DKIM both have the problem of ``identifier alignment''. It means that the sender email address that user sees can be different from the address that is actually used to do perform authentication. For SPF, the authentication focuses on the ``Return-Path'' and examines whether the sender's IP is listed in the ``Return-Path'' domain's SPF
record.  An attacker can set the ``Return-Path'' domain to her own domain and set her SPF record to pass the authentication. However, what the receiving user sees on the email interface is set by the ``From'' field. 
DKIM has a similar problem given that the domain to sign the email with the DKIM key can be different from the domain on the ``Return-Path'' .  DMARC helps to revolve the prolem by enforing the alignment of the identifiers. 

Second, mail forwarding is a problem for SPF. Mail forwarding means one email service automatically forwards emails to another email service. A common scenario is that university students often configure their university email service to forward all their emails to Outlook or Gmail. During Mail forwarding, the email metadata ({\em e.g.}, ``Return-Path'') remains unchanged. SPF will fail after mail forwarding because the forwarder's IP will not match the original sender's SPF record. 


Thrid, mailing list is a major problem for both SPF and DKIM. When a message is sent to a mailing list, the mailing list will ``broadcast'' the message to all the  subscribers. This is a similar process as mail forwarding. During this process, the mailing list's IP will become the sender IP, which is different from the original sender's IP. This will leads to SPF failure.  Mailing lists will cause trouble for DKIM because most mailing lists modify the email content before
broadcasting it to the subscribers. The common modification is to add a ``footer'' with the mailing list's name and a link for un-subscription. Tempering the email content will cause DKIM failure. DMARC helps to solve some of the problems, but not the mailing list problem.  For mailing lists, DMARC+SPF will be sure to fail --- if the ``Return-Path'' is modified, DMARC
will fail due to the misalignment of identifiers; if the ``Return-Path''  is unmodified, SPF will fail due to the IP mismatch. For DMARC+DKIM, it will fail if the mailing list still has to modify the email content. 

In  particular, {\em U7}pointed out the problem of DKIM beyond just the mailing list problem. {\em U7} stated that DKIM was too sensitive to ``benign'' changes to the email content such as line rewrapping and URL expansion. These operations that are very common in email services (sometimes for usability purposes), but can easily lead to invalid signatures. The sensitivity of DKIM also discourages email administrators to deploy DMARC (which need to work with DKIM). 

\begin{quote}
  {\em 
  ``U7: DKIM is inherently flawed because semantically meaningless changes to a message can render the signature invalid. 
  For example, the relaxed body canonicalization algorithm is sensitive to line rewrapping, 
  which will invalidate the signature without changing the semantic content of the message. 
  Flaws like this make DKIM signatures fragile,
  reducing the utility of DKIM and thus lessening the priority of its deployment.''
  }
  \end{quote}

  \begin{quote}
    {\em 
    ``U7: The fragility of DKIM also affects the utility of DMARC, and thus reducing the priority of its deployment as well.''
    }
  \end{quote}


\subsection{A Lack of Critical Mass}  
Email administrators mentioned that there had not been a global
consensus that SPF, DKIM or DMARC should be the ultimate solution to
stop spoofing. Part of the reason is these protocols are
struggling to support common email scenarios such as mail
forwarding. Due to the technique weaknesses, the general perception is that SFP, DKIM
and DMARC are ``helpful'' but ``cannot solve the spoofing problem
completely''. {\em U2} mentioned that potential adopters could be are waiting
to see whether enough people would eventually get on board. 

\begin{quote}
{\em ``U2: It is not the final answer that the industry picked up yet.
I felt at this point that enough people haven't really adopted it,
it's not worth for me to set it up.''
}
\end{quote}

In addition, {\em U1} and {\em U2} both mentioned that in general there was no penalty to
domains for not publishing an SPF/DKIM/DMARC record. Their emails are typically not discriminated unless other
malicious signals are detected. This reflects a typical bootstrapping challenge, where a ``critical mass'' is
needed in order to facilitate a self-sustaining adoption process~\cite{ozment2006boot}.

\subsection{Benefits Not Significantly Overweight Costs}  
Email administrators then discussed the deeper reasons for the lack of critical
mass. {\em U1} pointed out that the protocol adopter does not directly benefit from
publishing their SPF, DKIM or DMARC records in the DNS. Instead, these DNS
records mainly help {\em other email services} to verify incoming
emails and protect the customers (users) of other email
services. Domains that publish the DNS records receive the benefit of
a better reputation, which is a relatively vague benefit, particularly for
domains that don't host email services. 


\begin{quote}
{\em ``U1: If I am an email provider, I am not motivated to set up SPF,
I am motivated to make sure people who have sent (emails) to my customers have set SPF.
I am motivated to evaluate it.''}
\end{quote}

For popular online services ({\em e.g.}, social networks, banks),
however, they are likely to be motivated to
publish SPF, DKIM, and DMARC records to prevent being spoofed and maintain
their good reputation ({\em U2, U3}). 


\begin{figure}
\begin{center}
\begin{minipage}[t]{0.29\textwidth}
\includegraphics[width=0.99\textwidth]{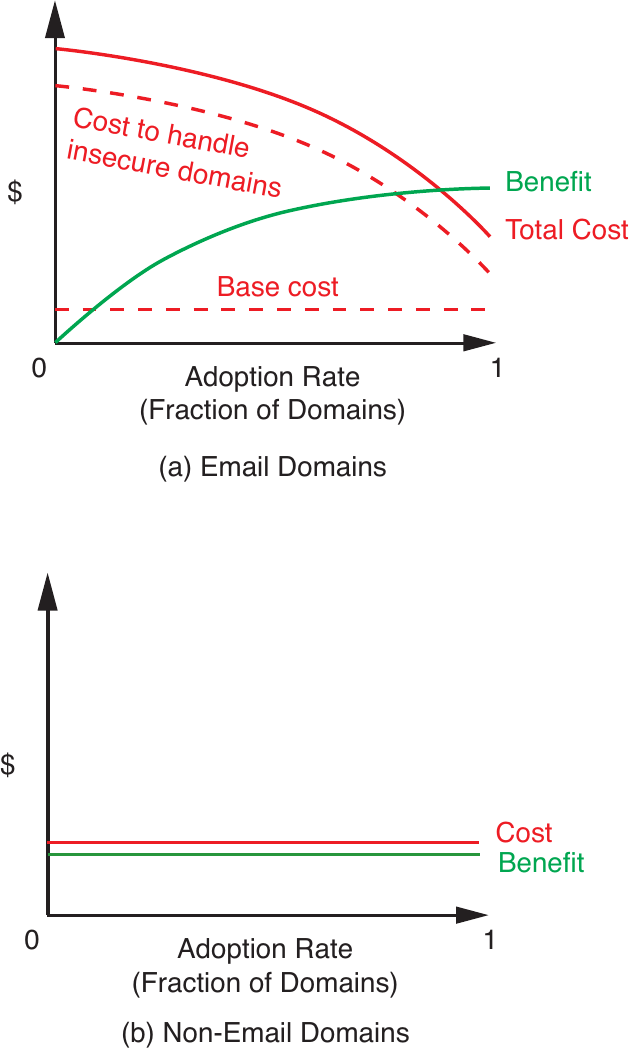}
\end{minipage}
\caption{The adoption model for anti-spoofing
  protocols. For email domains, the cost and benefit changes as more
  domains adopt the protocol. For non-email domains, the cost and
  benefit stay constant. }
\label{fig:model}
\end{center}
\end{figure}

To help to illustrate this challenge, we plot Figure~\ref{fig:model},
which is a modified version of the Ozment-Schechter
model~\cite{ozment2006boot}. Ozment-Schechter
model depicts the general challenge for network protocols to receive a
wide adoption, and we customized the model for email spoofing
scenarios and created a separate plot for non-email domains (Figure~\ref{fig:model}(b)). 

For email domains (Figure~\ref{fig:model}(a)), when more domains
publish their SPF, DKIM or DMARC records, the benefits for each
adopter will increase because more incoming emails can be
authenticated. Regarding the costs, there will be a constant {\em
  base cost} for deploying the protocol. On top of that, early
adopters also need to handle the insecure domains that have not
adopted the protocol and those with misconfigurations. The cost of insecure domains will drop as more
domains adopt the protocol. However, this cost cannot reach zero due to the
technical issues in these protocols as discussed before.

Figure~\ref{fig:model}(b) shows a bigger challenge to
motivate non-email domains to publish the SPF/DMARC record. 
For non-email domains ({\em e.g.}, {\tt
  office.com}), the benefit of publishing the SPF/DMARC record is to
prevent attackers from impersonating the non-email domain and helps the non-email domain to maintain a good
reputation. The domain administrators publish the SPF/DMARC
records to be a good Internet ``citizen'' and help other email
services to detect spoofing emails. However, these benefits are considered indirect and thus
relatively weaker ({\em U5, U6}). Overall, the cost and benefit model is not in favor of creating a
``critical mass'' for a wide adoption. The bootstrapping phase is
challenging without external enforcement or incentives. 


 
\subsection{Deployment Difficulties in Practice}
Even if an email administrator decided to deploy the protocol,
there would be other challenges in the way. We summarize the
participants' responses from three aspects: (1) a lack of control on the
DNS or even the mail servers, (2) the large number of dependency
services, (3) a lack of understanding of the protocol and the
deployment difficulties. 

First, certain services do not have a control over their DNS
record. Publishing SPF/DKIM/DMARC record will incur additional
overhead to coordinate with their DNS providers ({\em U1, U4, U9}).
In addition, many companies and organizations even don't
maintain their own mail servers but rely on cloud-based email
services. Using cloud-based email services is convenient without the
need the handle challenging tasks such as spam filtering. The drawback
is that the organization need to rely on the cloud email service to
deploy the anti-spoofing protocols. 

\begin{quote}
{\em
``U1: So we have very limited control over our DNS. Right now, it is
just the difficulty of setting up that DNS.''
}
\end{quote}


Another challenge is that the strict enforcement of certain email
protocols requires significant efforts for coordination in big institutions. An email system
has many dependent services ({\em e.g.}, marketing tools) distributed in different departments in a
big institution. Deploying a new email protocol requires a
non-trivial collaboration effort from different departments.  

  \begin{quote}
  {\em
  ``U7: Strict enforcement requires identifying all the legitimate sources of email using a return address domain. 
  Large, decentralized organizations (e.g. many large universities), 
  will often have organizational units which acquire third-party services involving email, 
  like email marketing tools, without telling central IT. 
  Figuring all this out and putting policies and procedures in place to prevent it is more work than many admins have time for.''
  }
  \end{quote}

Finally, the participants mentioned that there had been a lack of
deep understanding of the anti-spoofing protocols, especially the new protocols such as DMARC. It
is difficult to estimate how much effort is needed to deploy and
maintain the protocol in practice. {\em U3} particularly mentioned
that there is a general perception that deploying anti-spoofing
protocols is difficult. Regardless the actual level of the difficulty,
the perceived difficulty makes email administrators hesitated to try ({\em U3, U9}).  

  \begin{quote}
  {\em
  ``U3: Many people believe that DKIM is hard, and thus don't
  prioritize deploying it ... Many people don't understand DMARC, how easy it is to deploy, and how effective it is.''
  }
  \end{quote}

\subsection{Risks of Breaking the Existing System}
Participants have discussed the concerns of breaking the existing email
system due to unfamiliarity to the protocol. This is particularly true
for DMARC (published in 2015). Email providers need to go through
careful testing to make sure the protocol does not block legitimate
incoming emails, and their own emails are not blocked by others.

\begin{quote}
{\em 
``U2: Probably because it (DMARC) is still in a testing phase
and (people) want to see if it is going to work for them. Relatively it
(DMARC) is still pretty new for big businesses and such.''
}
\end{quote}

  \begin{quote}
  {\em 
  ``U5: Domains may fear that they've forgotten something and their email may be rejected due to a mistake on their part. ''
  }
  \end{quote}

These concerns also explain why most protocol adopters (as the sender
domain) configure a relaxed SPF/DMARC policy. Even if sender domain actually specified a strict protocol, the receiver may not enforce it anyway. {\em U5} expressed
that it was quite often for senders to have mis-configurations. It is
easier to not enforce the strict policy than to ask
the senders to fix their configurations.

  \begin{quote}
  {\em
  ``U5: Spam filters are relied upon too heavily and its sometimes easier to pull email from the spam folder 
  than ask someone to fix their SPF record and re-send the email.''
  }
  \end{quote}

\subsection{Solutions Moving Forward}
We asked the participants to comment on the possible solutions moving
forward. Most of the email administrators believed that automated detection
systems ({\em e.g.}, anti-spoofing protocols, spam filters, virus
scanners) were necessary, but could not fully prevent spoofing or phishing. 
{\em U1}, {\em U2}, {\em U7}, {\em U8} and {\em U9} all have
mentioned the importance of user education to raise
the awareness of spoofing, and training users to check the email authenticity
themselves. 
 

  \begin{quote}
  {\em 
  ``U7: There is no one single way. 
  Technological defenses like content filtering of incoming mail (i.e. spam and virus filtering), 
  are necessary but not sufficient. 
  There is also a need for rigorous training combined with periodic self-phishing (e.g. phishme.com), 
  to raise awareness and identify people who need further training or correction.''
  }
  \end{quote}

  \begin{quote}
  {\em 
  ``U8: User education is the most important way to protect them. 
    I always ask our users to look for the email that seems suspicious
    and bring it to my attention. That way we can prevent malicious intention at earliest possible.''
  }
  \end{quote}

Finally, {\em U5} expressed the need to have security indicators on
the email client. The security indicators are icons or visual cues that
are widely used on web browsers to indicate the validity of SSL
certificate of websites. A similar email spoofing indicator can be
deployed to warn users of emails with unverified sender addresses. In addition, security
indicators can also help to high-light the address misalignment of the {\tt
  Return-Path} and {\tt Mail From} fields for emails that bypassed the
SPF check. 

  \begin{quote}
  {\em 
  ``U5: Add the ability for email clients to warn users similar to the way 
  browsers do when users are either presented with a valid extended SSL cert or no SSL cert at all.  
  May also display the from \& reply to addresses making it harder to get around SPF record checking.''
  }
  \end{quote}

\section{Discussion}
So far, we have explored the challenges for SPF, DKIM and DMARC to
receive a wide adoption. Next, we discuss the key implications to
protocol designers, email providers, and the end users. 

\subsection{Implications for Protocol Designers and Promoters} 

\para{Improving the Perceived Usefulness.} The security and usability issues in SPF,
DKIM and DMARC negatively impact their perceived
usefulness. To improve the perceived usefulness, addressing these security and
usability issues becomes the first priority. Currently, an IETF group is
working on a new protocol called Authenticated Received
Chain (ARC)~\cite{arc} which is expected to address email forwarding
problem and the mailing list problem. However, this also adds to the number of protocols that domain
owners need to deploy. New protocols will have their own challenges to be
accepted. For example, the DMARC protocol, even though
incrementally deployable, only achieved a 4.6\% adoption rate in the
past two years. A useful protocol will still face the challenge to be
widely adopted.



\para{Building the Critical Mass. } Currently, there is a lack of
strong consensus to deploy anti-spoofing protocols. Like many
networking protocols, anti-spoofing protocols will provide key
benefits only after enough domains start to publish their SPF, DKIM or
DMARC records. To bootstrap the adoption and establish a critical mass, external
incentive mechanisms are needed. In theory, we can adjust the rewarding
function to provide more benefits to early adopters to create a
positive net effect~\cite{ozment2006boot}. One possible direction is
to learn from the promotion of  ``HTTPS'' among
websites~\cite{203662}: modern browsers will display a trusted icon
for websites with valid TLS certificates. Similar security indicators
can be added to emails with verified sender domains (by SPF, DKIM and DMARC), to incentive domains
to publish the corresponding DNS records. In addition, policymakers or major email
providers may also consider enforcing certain sensitive domains ({\em
e.g.}, banks, government agencies) to publish their SPF/DKIM/DMARC
records to prevent being impersonated. The challenge is how to realize
these ideas without disrupting any of the normal operations of the existing
email services. 

\para{Reducing the Deployment Difficulty. } One direction to improve the
adoption rate of anti-spoofing protocols is to make it easy to
deploy and configure. Our user study reveals two key problems to
address. First, more organizations start to use cloud-based email
services ({\em e.g.}, Google G-Suite, Amazon WorkMail, Office
365). Anti-spoofing protocols should be more cloud-friendly for
organizations that don't have full controls on their mail
servers. Second, the deployment process should be further
simplified and providers email administrators with more controls. The
biggest concern from email administrators is that anti-spoofing
protocols may reject legitimate emails or get their own emails
rejected. One direction of improvement is to allow the protocol to
run in a {\em testing mode} ({\em e.g.}, in DMARC), allowing email administrators to fully assess the
impact before real deployment. 

\subsection{Implications for Email Providers} 
In the short term, email providers are still unlikely to be able to authenticate {\em
all} the incoming emails. While email providers
should act as ``good Internet citizens'' by publishing their own
authentication records, it is also necessary to help to ``educate''
their users to watch out for spoofing emails. Given the current
adoption rate of anti-spoofing protocols (and the relaxed protocol
configurations), it is likely that email providers will
still have to deliver certain unverified emails to the user inbox. Email providers should act more responsibly by
providing the authentication results available for the user to check, or
proactively warn users of emails that they are not able to
verify. Large email providers such as Gmail and Outlook are already moving towards
this direction. Currently, Gmail's authentication results are
available through the webmail interface, but unfortunately not yet
available on the mobile app interface. Further research is needed to
improve the current mobile email UI to better support security
features.

\subsection{Implications for Users} 
Given the current situation, users are at the most vulnerable
position. Particularly, considering the usability flaws of the existing
anti-spoofing protocols, an email that passed
the SPF/DKIM checks can still be a spoofed email ({\em e.g.}, with misaligned
addresses). Similarly, emails that failed the SPF/DKIM checks are not
necessarily malicious ({\em e.g.}, forwarded email). To this end,
unless the user is fully aware of the authentication details, it is
safer for general email users to avoid establishing the trust based on the sender domains. The trustworthiness of
the email should be assessed as a whole. It is more reliable to
leverage the context of the email exchange, and the external
confirmation channels ({\em e.g.}, calling the sender on the phone) to identify phishing attempts and
securely handle critical emails.

\section{Limitations} 
The scale of the user study is still small, which limits us from producing any statistically significant results. We argue that our
contribution is to provide a ``qualitative'' understanding of the
problem space, which lays the groundwork for future quantitative
research. For example, one future direction is to conduct surveys to understand what types
of domains are more likely to adopt anti-spoofing protocols, and how
domain attributes ({\em e.g.}, service type, popularity, sensitivity)
affect the domain owners' decision. 


\section{Conclusion}
In this paper, we examine why email spoofing is (still) possible in today's email system. 
We show that extensive efforts are needed to address the technical issues in the protocol design and develop external enforcement (or incentives) to bootstrap the protocol adoption. In addition, improved user interfaces are needed for email systems to allow users to proactively check the email authentication results.

\balance
\bibliographystyle{IEEEtranS}
\bibliography{bibliography,astro,wang,anon,social,education,zhao}

\end{document}